\documentclass[usenatbib]{mn2e}
\usepackage{color}
\usepackage{graphicx,graphics}
\usepackage{amssymb,amsmath}
\usepackage{natbib}

\title[Statistical Properties of Diffuse Ly$\alpha$ Halos]
 {
 Statistical Properties of Diffuse Ly$\alpha$ Halos around Star-forming Galaxies at $z\sim 2$
}

\author[R. Momose et al.]
{
Rieko~Momose$^{1,2}$,
Masami~Ouchi$^{1,3}$,
Kimihiko~Nakajima$^{3,4,5}$,
Yoshiaki~Ono$^1$,
\newauthor
Takatoshi~Shibuya$^{1,6}$,
Kazuhiro~Shimasaku$^{4,7}$, 
Suraphong~Yuma$^{1,8}$,
\newauthor
Masao~Mori$^6$,
and
Masayuki~Umemura$^6$
\newauthor
\\
$^1$
Institute for Cosmic Ray Research, The University of Tokyo, 
5-1-5 Kashiwanoha, Kashiwa, Chiba 277-8582, Japan \\
$^2$
National Astronomical Observatory of Japan, 
2-21-1 Osawa Mitaka Tokyo, 181-8588 Japan \\
$^3$
Kavli Institute for the Physics and Mathematics of the Universe (WPI), The University of Tokyo, 
5-1-5 Kashiwanoha, Kashiwa, \\
Chiba 277-8583, Japan \\
$^4$
Department of Astronomy, Graduate School of Science, The University of Tokyo, 
7-3-1 Hongo, Bunkyo-ku, Tokyo 113-0033, Japan \\
$^5$
Observatoire de Gen\`{e}ve, Universit\'{e} de Gen\`{e}ve, 51 Ch. des
Maillettes, 1290 Versoix, Switzerland
Switzerland \\
$^6$
Center for Computational Sciences, The University of Tsukuba, 
1-1-1 Tennodai, Tsukuba, Ibaraki 305-8577, Japan \\
$^7$
Research Center for the Early Universe, Graduate School of Science, 
The University of Tokyo, Tokyo 113-0033, Japan \\
$^8$
Department of Physics, Faculty of Science, Mahidol University, Bangkok 10400, Thailand \\
}

\date{}

\pagerange{\pageref{firstpage}--\pageref{lastpage}} \pubyear{2015}

\begin{document}

\label{firstpage}

\maketitle

\begin{abstract}
{
We present statistical properties of diffuse Ly$\alpha$ halos (LAHs) around high-$z$ star-forming galaxies
with large Subaru samples of Ly$\alpha$ emitters (LAEs) at $z=2.2$. We make subsamples 
defined by the physical quantities of LAEs' central Ly$\alpha$ luminosities, UV magnitudes, 
Ly$\alpha$ equivalent widths, and UV slopes, and investigate LAHs' radial surface brightness (SB)
profiles and scale lengths $r_n$ as a function of these physical quantities.
We find that there exist prominent LAHs around LAEs with faint Ly$\alpha$ luminosities, bright UV luminosities, 
and small Ly$\alpha$ equivalent widths in cumulative radial Ly$\alpha$ SB profiles.
We confirm this trend with the anti-correlation between $r_n$ and Ly$\alpha$ luminosities
(equivalent widths) based on the Spearman's rank correlation coefficient
that is $\rho=-0.9$ ($-0.7$) corresponding to the $96\%$ ($93\%$) confidence level,
although the correlation between $r_n$ and UV magnitudes is not clearly found in the rank correlation coefficient.
Our results suggest that LAEs with properties similar to typical Lyman-break galaxies
(with faint Ly$\alpha$ luminosities and small equivalent widths) possess more prominent LAHs.
We investigate scenarios for the major physical origins of LAHs with our results,
and find that the cold stream scenario is not preferred, due to the relatively small equivalent widths
up to $77$\AA\ in LAHs that include LAEs' central components. There remain 
two possible scenarios of Ly$\alpha$ scattering in circum-galactic medium
and satellite galaxies that cannot be tested with our observational data.
}
\end{abstract}


\section{Introduction}
\label{introduction}
Recent observations have identified diffuse Ly$\alpha$ halos (LAHs) around Ly$\alpha$ emitters (LAEs) and Lyman-break galaxies (LBGs) by the stacking analysis that enables to find very diffuse and faint Ly$\alpha$ emission associated with high-redshift galaxies \citep{haya04,stei11,matsu12,feld13,momo14}. \citet{stei11} have identified extended LAHs with a radius of $r\sim80$ kpc around LBGs at $\langle z\rangle=2.65$ in the stacked narrow-band (NB) images of $92$ LBGs. 
\citet[][hereafter MA12]{matsu12} have detected LAHs from stacked $130-864$ LAEs at $z=3.1$. 
More recently, \citet[][hereafter MO14]{momo14} have found LAHs 
based on the large samples of $100-3500$ LAEs at the wide redshift range, $z=2-7$.
These previous studies suggest the inherent presence of LAHs around LAEs and LBGs, whereas they provoke new questions: what is the origin of LAHs, and which physical parameters determine the structure of LAHs.

Theoretically, the scattered light of Ly$\alpha$ photons by neutral hydrogen (H\,{\sc i}) gas in the circum-galactic medium (CGM) is thought as one possible origin of an LAH (e.g. \citealp{lau07,zhen11,dijk12,verha12}). Moreover, these studies have predicted that a galactic scale outflow and the environment of galaxies could produce the extended feature of Ly$\alpha$ emission. 
\citet{zhen11} have investigated properties of LAHs based on Ly$\alpha$ radiative transfer modeling in cosmological reionization simulations, and indicated that the slope of a radial surface brightness (SB) profile depends on an outflowing velocity of the CGM. 
\citet{lake15} have explored the nature of Ly$\alpha$ photons in an LAH found in MO14 by comparing the LAH obtained from their cosmological hydrodynamic galaxy formation simulation. Their results have suggested that the extent of their LAH over $20$ kpc radius is contributed by Ly$\alpha$ photons from satellite galaxies and the cold streams. 
\citet{jee12} have demonstrated that the radial SB profiles of LAHs are flatter at the epoch of reionization than those at the post-reionization epoch, since Ly$\alpha$ photons scattered by H\,{\sc i} gas in the intergalactic medium (IGM).

Observationally, the correlation between the LAH sizes and the central galaxy properties has been investigated. MA12 have explored the scale lengths of their LAHs $r_n$ as a function of the surface number density of LAEs $\delta_{\text{LAE}}$, and found that scale lengths of their LAHs correlate with $\delta_{\text{LAE}}$. MA12 have also studied the dependence between scale lengths and UV-continuum magnitude of their LAEs, but no correlation has been identified. 
The dependence of scale lengths on UV luminosity has been discussed in \citet{stei11} and \citet{feld13} as well. \citet{stei11} have shown a marginal difference in scale lengths between UV luminous and faint LBG samples. On the other hand, \citet{feld13} have suggested that their LAHs of UV-bright galaxies are more extended than that of all galaxies. 
The discrepancy of these results is unclear. 
However, this would be explained by the following two reasons.
(1) Galaxies samples used in both MA12 and \citet{stei11} are found in over-dense regions, and their data include environmental effects. (2) A marginal detection of the LAH found in \citet{feld13} may not allow to measure its scale length accurately. 
In order to clearly understand the properties and physical origin of LAHs, the detailed LAH observations of galaxies in low-density environments are necessary.

In this paper, we investigate the LAH profiles and sizes as a function of observational properties of the central LAEs. Our LAE samples, which reside in field regions at $z=2.2$ (\citealp{naka12}, MO14), are large enough to detect LAHs of subsamples defined by various physical properties of the central LAEs. We show the sample and analysis in Section \ref{data_and_analysis}, and our results in Section \ref{sec:results}. In Section \ref{discussions}, we discuss the physical origin of LAHs and evaluate the missing Ly$\alpha$ fluxes. We summarize our results and discussions in Section \ref{summary}. Throughout this paper, we use AB magnitudes \citep{oke83} and adopt a cosmology parameter set of ($\Omega_{\text{m}}$, $\Omega_\Lambda$, $H_0$)$=$($0.3$, $0.7$, $70$ km s$^{-1}$ Mpc$^{-1}$). In this cosmology, $1$ arcsec corresponds to transverse sizes of $8.3$ physical kpc at $z=2.2$.

\begin{table}
\centering
\begin{minipage}{85mm}
\begin{center}
\caption{Definitions and Properties of our Subsamples}
\label{tab:sample}
\begin{tabular}{cccccc}
\hline
Subsample 				& Threshold 									& $N$ 		& $C_n$	& $r_n$	\\
(1)						& (2)											& (3)			& (4)		& (5)	\\
\hline
$\log{L_{\text{Ly$\alpha$}}}=42.6$	& $42.4\le\log{L_{\text{Ly$\alpha$}}} \ \ \ \ \ \ \ \ \ $	& 710		& $2.9$		& $7.3_{-0.30}^{+0.33}$\\
$\ \ \ \ \ \ \ \ \ \ \ \ \ \ \ 42.3$ 		& $42.2\le\log{L_{\text{Ly$\alpha$}}}<42.4$			& 711		& $1.5$		& $7.5_{-1.14}^{+1.64}$\\
$\ \ \ \ \ \ \ \ \ \ \ \ \ \ \ 42.1$ 		& $42.0\le\log{L_{\text{Ly$\alpha$}}}<42.2$			& 711		& $0.7$		& $10.5_{-1.07}^{+1.35}$	\\
$\ \ \ \ \ \ \ \ \ \ \ \ \ \ \ 41.9$ 		& $41.8\le\log{L_{\text{Ly$\alpha$}}}<42.0$			& 711		& $0.3$		& $16.2_{-3.49}^{+6.14}$	\\
$\ \ \ \ \ \ \ \ \ \ \ \ \ \ \ 41.7$ 		& $\ \ \ \ \ \ \ \ \ \log{L_{\text{Ly$\alpha$}}}<41.8$		& 711		& $0.5$		& $14.3_{-1.92}^{+2.28}$	\\
\hline
$M_{\text{UV}}=-21.1$					& $\ \ \ \ \ \ \ \ \ \ \ M_{\text{UV}}<-20.7$				& 710		& $1.8$		& $8.1_{-0.52}^{+0.51}$	\\
$\ \ \ \ \ \ \ \ \ \ \ -20.5$ 				& $-20.7\le M_{\text{UV}}<-20.2$						& 710		& $0.7$		& $9.0_{-0.96}^{+1.35}$	\\
$\ \ \ \ \ \ \ \ \ \ \ -20.1$ 				& $-20.2\le M_{\text{UV}}<-19.9$						& 710		& $0.9$		& $7.8_{-1.02}^{+1.49}$	\\
$\ \ \ \ \ \ \ \ \ \ \ -19.7$ 				& $-19.9\le M_{\text{UV}}<-19.4$						& 710		& $0.6$		& $9.9_{-2.46}^{+4.93}$	\\
$\ \ \ \ \ \ \ \ \ \ \ -18.9^\dagger$ 		& $-19.4\le M_{\text{UV}} \ \ \ \ \ \ \ \ \ \ \ $			& 710		& $0.7$ 	& $12.7_{-2.39}^{+3.49}$\\
\hline
$EW_0=150$							& $90\le EW_0 \ \ \ \ \ \ \ $							& 711		& $1.7$		& $8.3_{-0.47}^{+0.44}$	\\
$\ \ \ \ \ \ \ \ \ \ \ \ 63$ 				& $49\le EW_0<90$									& 711		& $1.1$		& $9.7_{-0.64}^{+0.61}$	\\
$\ \ \ \ \ \ \ \ \ \ \ \ 40$ 				& $34\le EW_0<49$									& 711		& $0.7$		& $10.2_{-0.92}^{+0.97}$	\\
$\ \ \ \ \ \ \ \ \ \ \ \ 30$ 				& $26\le EW_0<34$									& 711		& $0.4$		& $14.5_{-1.59}^{+2.03}$	\\
$\ \ \ \ \ \ \ \ \ \ \ \ 22$ 				& $\ \ \ \ \ \ \ EW_0<26$								& 712		& $0.4$		& $10.5_{-5.17}^{+72.28}$	\\
\hline
$\beta=0.7$							& $-0.02\le\beta \ \ \ \ \ \ \ \ \ \ \ $					& 659		& $0.3$		& $12.0_{-1.90}^{+2.52}$	\\
$\ \ \ \ \ \ \ -0.5$ 						& $-1.02\le\beta<\ \ 0.02$								& 659		& $0.5$		& $11.3_{-1.12}^{+1.59}$	\\
$\ \ \ \ \ \ \ -1.4$ 						& $-1.72\le\beta<-1.02$								& 659		& $0.8$		& $9.1_{-0.90}^{+1.12}$	\\
$\ \ \ \ \ \ \ -2.0$ 						& $-2.25\le\beta<-1.72$								& 659		& $1.3$		& $8.9_{-0.54}^{+0.61}$	\\
$\ \ \ \ \ \ \ -2.6$ 						& $\ \ \ \ \ \ \ \ \ \ \ \beta<-2.25$						& 660		& $1.1$		& $9.5_{-0.61}^{+0.70}$	\\
\hline
$\delta_{\text{LAE}}=0.73^{\dagger\dagger}$	& $0.5\le\delta_{\text{LAE}}<1.5$ 				& 1047		& $2.2$ & $7.4_{-0.61}^{+0.72}$\\
$\ \ \ \ \ \ \ \ \ \ \ 0.04^{\dagger\dagger}$ & $-1\le\delta_{\text{LAE}}<0.5$ 					& 348		& $1.9$ & $7.7_{-0.35}^{+0.38}$\\
\hline
\end{tabular}
\end{center}
(1) Subsample name that indicates the median value of $\log{L_{\text{Ly$\alpha$}}}$, $M_{\text{UV}}$, $EW_0$, $\beta$, or $\sigma_{\text{LAE}}$;
(2) threshold of the subsample. The $\log{L_{\text{Ly$\alpha$}}}$ and $EW_0$ values are shown in units of erg s$^{-1}$ and \AA, respectively; 
(3) number of LAEs in the subsample; 
(4) best-fit $C_n$ in units of $10^{-18}$ erg s$^{-1}$ cm$^{-2}$ arcsec$^{-2}$;
(5) best-fit $r_n$ in units of kpc.
$^\dagger$: The subsample includes 54 LAEs with no detectable continuum in our $V$-band images.
$^{\dagger\dagger}$: 
These two subsamples have the Ly$\alpha$ luminosity threshold of 
$>1.5\times10^{42}$ erg s$^{-1}$
same as the one of \citet{matsu12}. 
\end{minipage}
\end{table}

\begin{table*}
\centering
\begin{minipage}{168mm}
\begin{center}
\caption{Results of the Kolmogorov-Smirnov Tests}
\label{tab:ks}
\begin{tabular}{cccccc}
\hline
(1)										& (2)				& (3)					& (4)				& (5)	\\
Name									& $\log{L_{\text{Ly$\alpha$}}}$	& $M_{\text{UV}}$	& $EW_0$	& $\beta$	\\
										& ($D$, $p$-value)	& ($D$, $p$-value)	& ($D$, $p$-value)	& ($D$, $p$-value)	\\
\hline
$\log{L_{\text{Ly$\alpha$}}}=42.6$	& --					& --					& --					& --	\\
$\ \ \ \ \ \ \ \ \ \ \ \ \ \ \ 42.3$ 		& --					& (0.23, 1.75e-17)	& (0.21, 3.61e-14)	& (0.07, 0.07)	\\
$\ \ \ \ \ \ \ \ \ \ \ \ \ \ \ 42.1$ 		& --					& (0.32, 3.60e-33)	& (0.36, 5.42e-41)	& (0.13, 8.86e-6)	\\ 
$\ \ \ \ \ \ \ \ \ \ \ \ \ \ \ 41.9$ 		& --					& (0.41, 3.33e-53)	& (0.49, 2.17e-75)	& (0.20, 2.19e-12)	\\ 
$\ \ \ \ \ \ \ \ \ \ \ \ \ \ \ 41.7$ 		& --					& (0.51, 3.13e-83)	& (0.56, 9.41e-100)& (0.26, 4.97e-21)	\\ 
\hline
$M_{\text{UV}}=-21.1$					& --					& --					& --					& --	\\			
$\ \ \ \ \ \ \ \ \ \ \ -20.5$ 				& (0.37, 2.76e-44)	& --					& (0.02, 0.99)		& (0.08, 0.03)	\\
$\ \ \ \ \ \ \ \ \ \ \ -20.1$ 				& (0.46, 1.60e-67)	& --					& (0.09, 3.30e-3)	& (0.10, 8.06e-4)	\\
$\ \ \ \ \ \ \ \ \ \ \ -19.7$ 				& (0.49, 1.70e-75)	& --					& (0.20, 1.85e-13)	& (0.14, 7.11e-7)	\\
$\ \ \ \ \ \ \ \ \ \ \ -18.9$ 				& (0.43, 3.14e-59) 	& --					& (0.55, 3.41e-95)	& (0.33, 1.67e-33)	\\
\hline
$EW_0=150$							& --						& --					& --					& --	\\
$\ \ \ \ \ \ \ \ \ \ \ \ 63$ 				& (0.13, 1.94e-5)		& (0.38, 6.93e-46)	& --					& (0.11, 6.59e-4)	\\
$\ \ \ \ \ \ \ \ \ \ \ \ 40$ 				&(0.27, 7.32e-24)		& (0.49, 7.98e-75)	& --					& (0.17, 2.61e-9)	\\
$\ \ \ \ \ \ \ \ \ \ \ \ 30$ 				&(0.41, 1.00e-52)		& (0.53, 8.30e-88)	& --					& (0.20, 7.21e-13)	\\
$\ \ \ \ \ \ \ \ \ \ \ \ 22$ 				& (0.55, 2.14e-95)		& (0.56, 7.78e-100) & --				& (0.27, 6.79e-24)	\\
\hline
$\beta=0.7$							& --						& --					& --					& --	\\
$\ \ \ \ \ \ \ -0.5$ 						& (0.10, 4.70e-3)		& (0.28, 6.40e-23)	& (0.07, 0.07)		& --	\\
$\ \ \ \ \ \ \ -1.4$ 						& (0.10, 4.70e-3)		& (0.28, 6.40e-23)	& (0.07, 0.07)		& --	\\
$\ \ \ \ \ \ \ -2.0$ 						& (0.28, 6.08e-24)		& (0.40, 1.67e-47)	& (0.11, 4.81e-4)	& --	\\
$\ \ \ \ \ \ \ -2.6$ 						& (0.26, 2.76e-20)		& (0.28, 1.88e-23)	& (0.25, 1.05e-18)	& --	\\
\hline
\end{tabular}
\end{center}
(1) Subsample name that indicates the median value of $\log{L_{\text{Ly$\alpha$}}}$, $M_{\text{UV}}$, $EW_0$, $\beta$, or $\sigma_{\text{LAE}}$;
(2) K-S statics $D$ and two-tailed $p$-value for the histograms of $\log{L_{\text{Ly$\alpha$}}}$ .
The K-S statistics is performed with the reference subsamples of $\log{L_{\text{Ly$\alpha$}}}=42.6$, $M_{\text{UV}}=-21.1$, $EW_0=150$, 
and $\beta=0.7$ for 
the subsamples of $\log{L_{\text{Ly$\alpha$}}}$, $M_{\text{UV}}$, $EW_0$, and $\beta$, respectively;
(3) same as (2), but for the histograms of $M_{\text{UV}}$; 
(4) same as (2), but for the histograms of $EW_0$; 
(5) same as (2), but for the histograms of $\beta$. 
\end{minipage}
\end{table*}


\section{Sample and Analysis}
\label{data_and_analysis}
\subsection{Sample}
We use a large photometric sample of LAEs at $z=2.2$ made by the wide-field narrow-band (NB) imaging surveys of Suprime-Cam \citep{miya02} on Subaru telescope. 
Our sample consists of $3556$ LAEs found in five deep fields of COSMOS, GOODS-N, GOODS-S, SSA22 and SXDS \citep{naka12}. The LAEs are identified by an excess of flux in an NB of $NB387$ whose central wavelength and FWHM are $3870$ {\AA} and $94$ {\AA}, respectively. The UV continua of these LAEs are determined with $V$-band images taken by \citet{cap04}, \citet{haya04}, \citet{tani07}, \citet{furu08}, and \citet{tayl09}.

To find a correlation between the LAH profiles and LAE properties, we divide our LAEs into five subsamples based on Ly$\alpha$ luminosity $\log{L_{\text{Ly$\alpha$}}}$, absolute continuum magnitude $M_{\text{UV}}$, Ly$\alpha$ rest-frame equivalent width $EW_0$, and UV spectral slope $\beta$. 
These observational quantities of LAEs except for $\beta$ are evaluated in the same manner as \citet{naka12}.
The $EW_0$ values are estimated from the observed $u^*$$- NB387$ and/or $B-NB387$ colors of LAEs. The $\log{L_{\text{Ly$\alpha$}}}$ is calculated from the $EW_0$ and total $u^*$ and/or $B-$band magnitudes. We regard observed $V$-band magnitudes as $M_{\text{UV}}$.
The $\beta$ values are estimated by the weighted least square fitting to three data points of $V$, $R$, and $i'$-band magnitudes. A $1\sigma$ limiting magnitude is used for the weight of the data point. We omit LAEs which are not detected in the three bands of $V$, $R$, and $i'$ band.
Using each quantity of $\log{L_{\text{Ly$\alpha$}}}$, $M_{\text{UV}}$, $EW_0$ or $\beta$, we divide our LAE sample into five subsamples which include the same number of LAEs. The thresholds of the quantities are presented in Table \ref{tab:sample}. To improve the signal-to-noise ratio of composite images, we remove some LAEs near bright Galactic stars from our subsamples.
Finally, each subsample consists of about 700 LAEs.
We derive median values of these subsamples 
summarized in Table \ref{tab:sample}.
In the following section, we refer to these subsamples as the median values,
for examples,
$\log{L_{\text{Ly$\alpha$}}}=42.6$, $42.3$, $42.1$, $41.9$, and $41.7$ for the five Ly$\alpha$ luminosity subsamples.
We also examine the $r_n$ values of our LAHs as a function of the LAE surface density, $\delta_{\text{LAE}}$, for comparing results of MA12. 
See Section \ref{sec:sd} for the definition of our $\delta_{\text{LAE}}$ subsamples.

\begin{figure}
	\begin{center} 
		\includegraphics[scale=0.14]{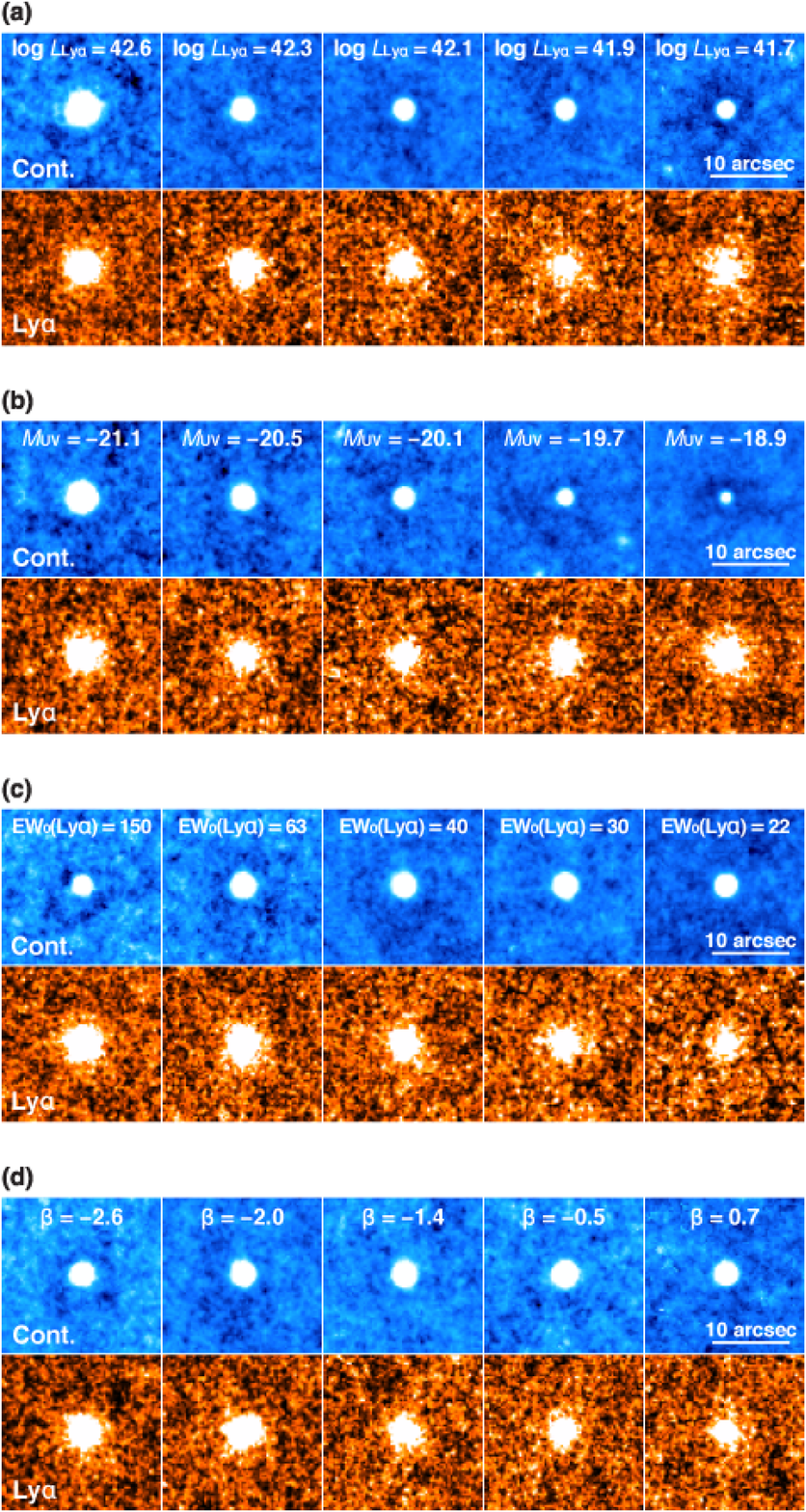}
	\end{center}
	\caption{Composite images of our LAE subsamples. The panel sets of (a)-(d) show
	 $\log{L_{\text{Ly$\alpha$}}}$-, $M_{\text{UV}}$-, $EW_0$-, and $\beta$-subsample composites, respectively. 
	 In each set of the panels, the UV-continuum and Ly$\alpha$ emission images are presented in the top and bottom rows, respectively.
	}
	\label{fig:image}
\end{figure}

\subsection{Properties of the Subsamples}
In order to assess statistical differences in physical properties of our subsamples, we carry out Kolmogorov-Smirnov (K-S) tests, and obtain $D$ and $p$ values between two of any subsamples in the $\log{L_{\text{Ly$\alpha$}}}$, $M_{\text{UV}}$, $EW_0$ and $\beta$ axes.
The K-S tests determine whether two subsamples are drawn from the same continuous distribution. 
There exists a significant difference between two samples, if the $p$-value is below $1\%$. 
The K-S test is performed between the reference $\log{L_{\text{Ly$\alpha$}}}=42.6$ subsample and the rest of four $\log{L_{\text{Ly$\alpha$}}}$ subsamples. 
Similarly, we conduct K-S tests for $M_{\text{UV}}$, $EW_0$, and $\beta$ subsamples with the references of $M_{\text{UV}}=-21.1$, $EW_0=150$, and $\beta=0.7$ subsamples, respectively. 
Table \ref{tab:ks} presents the $D$ and $p$ values of the K-S tests.
In the five out of forty eight K-S test results, $p$-values are above $1$\% (e.g. $\log{L_{\text{Ly$\alpha$}}}=42.6$ and $42.3$ subsamples in the $\beta$ distribution).
However, the majority, the rest of forty three, K-S test results show $p$-values less than $1$\%.
In other words, we rule out the null hypothesis that the majority of subsamples are drawn from the same sample.
From these results, we clarify that the most of subsamples depend not only on the quantities of the subsample definitions,
but also on the other quantities.

\subsection{Image stacking}
\label{image_stacking}
\begin{figure}
	\begin{center}
		\includegraphics[scale=0.25]{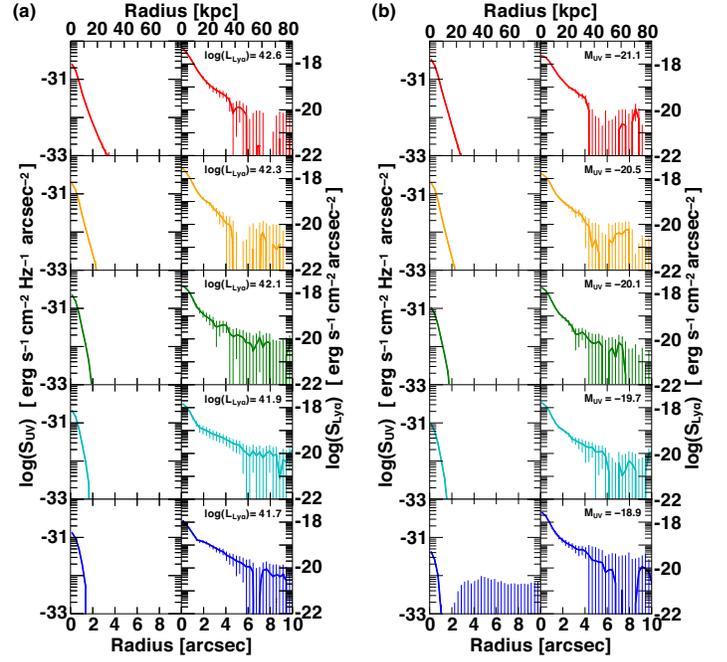} 
	\end{center}
	\caption{Differential radial SB profiles of our LAE subsamples. The panel sets of (a) and (b) represent the SB profiles of 
	$\log{L_{\text{Ly$\alpha$}}}$ and $M_{\text{UV}}$ subsamples, respectively. In each panel set, the left and right panels show 
	the SB profiles of UV-continuum and Ly$\alpha$ emission, respectively, with the colored lines. 
	The UV-continuum profiles are under the influence of sky over-subtraction systematics at the SB profile levels 
	below $\sim 4$ $\times$ $10^{-33}$ erg s$^{-1}$ cm$^{-2}$ Hz$^{-1}$ arcsec$^{-2}$ (see text for more details).
	}
	\label{fig:radi_all_1}
\end{figure}

\begin{figure}
	\begin{center}
		\includegraphics[scale=0.25]{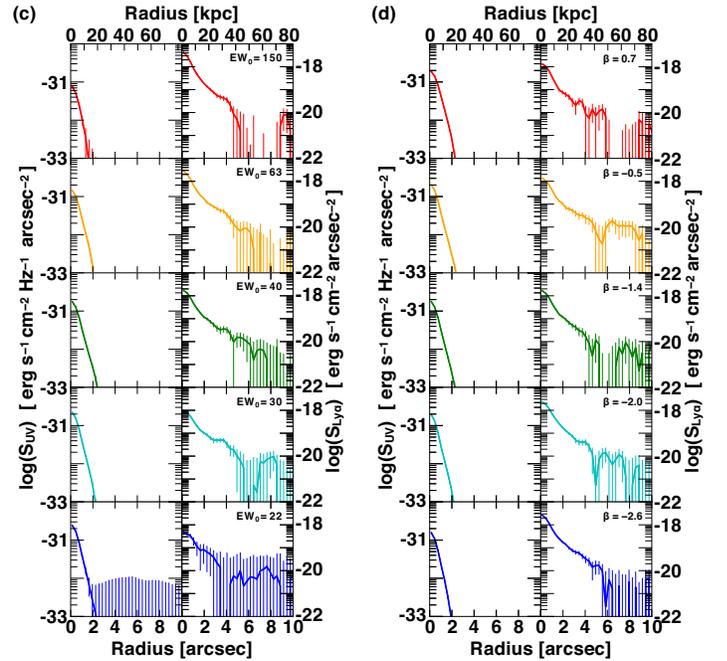} 
	\end{center}
	\caption{Same as Figure \ref{fig:radi_all_1}, but for the $EW_0$ (c) and $\beta$ (d) subsamples. 
	}
	\label{fig:radi_all_2}
\end{figure}

To investigate LAHs, we carry out stacking analysis with the UV-continnum ($V$-band) and NB images of our LAEs. We adopt a weighted-mean algorithm with a 1$\sigma$ error defined in each survey field. We follow the procedure of MO14 to make composite images. 
We subtract the composite UV-continuum images from NB images to obtain Ly$\alpha$ images.
In the same manner as MO14, we make sky images that are composite images with no objects within a $45'' \times 45''$ area, and produce the PSF images that are stacked images of point sources.
The composite images are presented in Figure \ref{fig:image}. 
We derive radial SB profiles from $r=0$ to $10$ arcsec with an annulus of $0.3$-arcsec width,
and present the differential radial SB profiles measured from the UV-continuum and Ly$\alpha$ images in Figures \ref{fig:radi_all_1} and \ref{fig:radi_all_2}, respectively.

\subsection{Systematic uncertainties}
\label{nonLAE_stacking}
We find some residuals of sky subtraction in the composite UV-continuum images and their differential radial profiles, due to the sky over-subtraction. 
Similarly, the same level of sky over-subtraction would exist in the NB images. 
Because the Ly$\alpha$ images are the NB images with the UV-continuum image subtractions,
it is likely that the sky over-subtraction effects mostly cancel out for the Ly$\alpha$ images. 
Nevertheless, we evaluate the impact on the UV-continuum profile from the sky over-subtraction effects.
We find that the UV-continuum profile has the effects of the sky over-subtraction at the level of $\lesssim4.3$ $\times$ $10^{-33}$ erg s$^{-1}$ cm$^{-2}$ Hz$^{-1}$ arcsec$^{-2}$ that corresponds to the maximum negative value of the profiles.

\begin{figure}
	\begin{center}
		\includegraphics[scale=0.125]{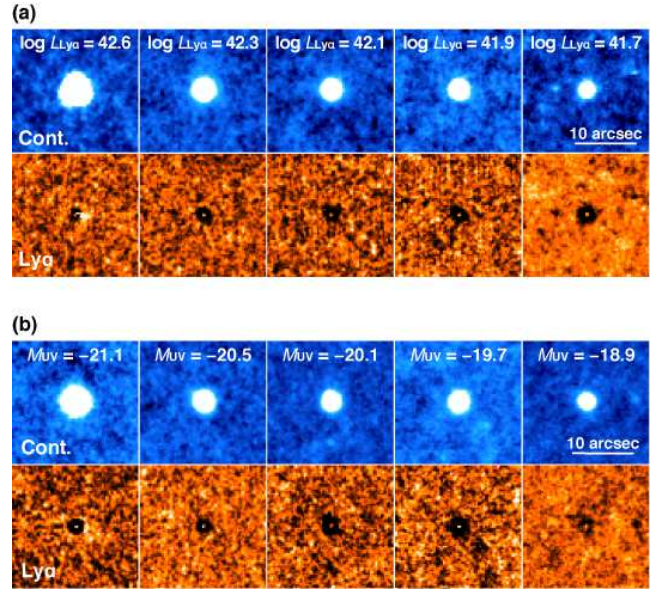} 
	\end{center}
	\caption{
	Same as Figure \ref{fig:image}, but for the non-LAEs that mock the LAE $\log{L_{\text{Ly$\alpha$}}}$ and $M_{\text{UV}}$ subsamples.
		}
	\label{fig:image_non}
\end{figure}

\begin{figure}
	\begin{center}
		\includegraphics[scale=0.4]{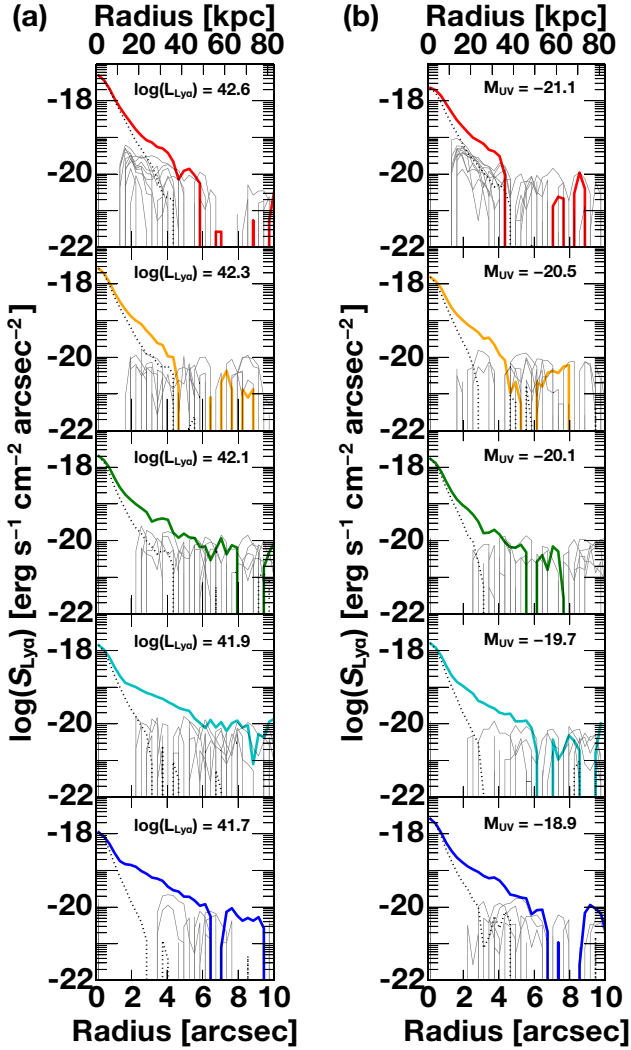} 
	\end{center}
	\caption{Differential radial Ly$\alpha$ SB profiles of LAEs (solid lines), non-LAEs (gray solid lines), and PSFs (dotted lines). 
	The panel sets of (a) and (b) represent the SB profiles for the $\log{L_{\text{Ly$\alpha$}}}$ and $M_{\text{UV}}$ subsamples, respectively.
	The gray solid lines represent the $10$ realizations of the non-LAE SB profile measurements. 
	}
	\label{fig:nonradi}
\end{figure}

We further evaluate the total systematic uncertainties including these sky over-subtraction effects by carrying out image stacking for objects that are not LAEs (see MO14).
Hereafter, these objects are referred to as non-LAEs.
We randomly choose non-LAEs with the number and magnitude distributions same as our LAE subsamples.
We perform image stacking for the non-LAEs in the same manner as Section \ref{image_stacking}.
We repeat this process 10 times to evaluate uncertainties of these non-LAE estimates. 
Figure \ref{fig:image_non} represents composite UV-continuum and Ly$\alpha$ images of non-LAEs that correspond to the LAE $\log{L_{\text{Ly$\alpha$}}}$ and $M_{\text{UV}}$ subsamples. 
Here, for simplicity we refer to the composite $V$-band images (the composite $V$-band images subtracted from $NB$-band images) as the UV-continuum (Ly$\alpha$) images of non-LAEs, although the $V$ and $NB$ bands mostly samples neither the rest-frame UV-continuum nor Ly$\alpha$ of non-LAEs whose redshifts are unknown.
In Figure \ref{fig:image_non}, we identify no significant extended profiles in the Ly$\alpha$ images of the non-LAEs. 
In the Ly$\alpha$ images, there exist ring-like structures near the source centers.
These structures are made by the slight differences between PSF profiles of $V$-band and NB images, which provide negligible effects on the evaluation of large-scale extended profiles.
Figure \ref{fig:nonradi} shows radial SB profiles of these UV-continuum and Ly$\alpha$ images of non-LAEs, together with those of the corresponding LAE subsamples.
In Figure \ref{fig:nonradi}, we find artificial extended profiles at the level of $10^{-20}$ erg s$^{-1}$ cm$^{-2}$ arcsec$^{-2}$ for all subsamples. However, these artifacts are significantly smaller than those of the LAE subsamples at $\lesssim 4-5''$.
Thus, we have identified LAHs in our subsamples that are produced by neither the statistical nor systematic uncertainties.
Moreover, these tests confirm that the total of systematic uncertainties produce spatially extended profiles only at the Ly$\alpha$ SB level of $\lesssim 10^{-20}$ erg s$^{-1}$ cm$^{-2}$ arcsec$^{-2}$.
In our analysis we conservatively use the Ly$\alpha$ profiles above the level of $\simeq$ $10^{-20}$ erg s$^{-1}$ cm$^{-2}$ arcsec$^{-2}$.

\subsection{Comparisons with the previous study}
\label{sec:sd}
\begin{figure}
	\begin{center}
		\includegraphics[scale=0.45]{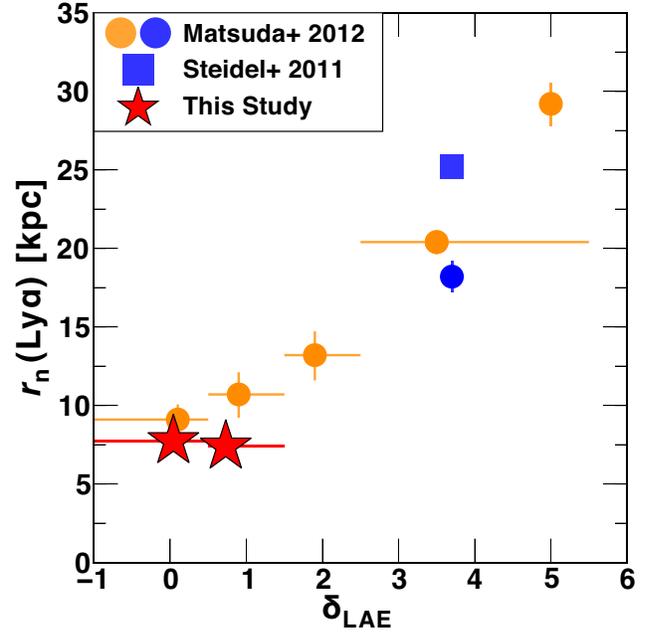} 
	\end{center}
	\caption{Scale length, $r_n$, as a function of $\delta_{\text{LAE}}$. The red stars represent our measurements. The orange circles denote the $r_n$ values of LAEs reported in MA12. The blue square and circle indicate the $r_n$ values of LBGs shown in Steidel et al. (2011) and MA12, respectively. 
	The vertical error bars are $1\sigma$ uncertainties of $r_n$ estimates, while horizontal error bars are the $\delta_{\text{LAE}}$ ranges of each subsample.
	}
	\label{fig:size_sd}
\end{figure}

We compare the LAHs' SB profiles of our LAE subsamples with those of MA12.
Because the sample definitions of ours and MA12 are different,
we make another set of LAE subsamples.
First, we select LAEs brighter than $1.5\times10^{42}$ erg s$^{-1}$ which is the same Ly$\alpha$ luminosity limit as that of MA12 \citep{yam12}. Second, we place another criterion of the LAE surface density, $\delta_{\text{LAE}} \equiv (\Sigma - \bar{\Sigma})/\bar{\Sigma}$, where $\Sigma$ is a surface density of LAEs within a radius of $10$ arcmin, and $\bar{\Sigma}$ is the field average of $\Sigma$.
MA12 use the criteria of $-1<\delta_{\text{LAE}}<0.5$, $0.5<\delta_{\text{LAE}}<1.5$, $1.5<\delta_{\text{LAE}}<2.5$ and $2.5<\delta_{\text{LAE}}<5.5$,
and accordingly we select 1047, 348, 0 and 0 LAEs, respectively, for these criteria.
We obtain only two LAE subsamples of 1047 and 348 LAEs with $-1<\delta_{\text{LAE}}<0.5$ and $0.5<\delta_{\text{LAE}}<1.5$, that are referred to as $\delta_{\text{LAE}}=0.04$ and $\delta_{\text{LAE}}=0.73$ subsamples, respectively.
We find that our LAEs reside in an environment whose LAE density is lower than that of MA12 who investigate the high LAE density region of the SSA22 proto-cluster.
We conduct image stacking of our $\delta_{\text{LAE}}=0.04$ and $\delta_{\text{LAE}}=0.73$ subsamples in the same manner as Section \ref{image_stacking}, and obtain the UV-continuum and Ly$\alpha$ images. The differential radial SB profiles are derived, and the LAHs are found.
To compare the SB profiles of ours with those of MA12, we perform the profile fitting 
to the SB profiles of the LAHs.
Following the previous work of MA12, we use the exponential profile defined by
\begin{equation}\label{eq:exp}
	S(r) = C_n \exp{(-r/r_n)},
\end{equation}
where $S(r)$, $r$, $C_n$, and $r_n$ are the differential radial SB profile, radius, normalization factor, and scale length, respectively. 
For our LAE $\delta_{\text{LAE}}=0.73$ and $0.04$ subsamples, we carry out the profile fitting to the LAHs in a radius range from $r=2''$ to $40$ kpc that is the same as our previous work (MO14). 
This radius range allows us to obtain $r_n$ with negligible contaminations of PSF ($r\sim0.6''$) and reasonably high statistical accuracies, avoiding the radius range under the influence of systematics ($r\gtrsim 5''$; Section \ref{nonLAE_stacking}).
The $r_n$ values of our $\delta_{\text{LAE}}$ subsamples are presented in Figure \ref{fig:size_sd}.
We find that $r_n$ measurements of our study are consistent with those of MA12.

\section{Results}
\label{sec:results}
In this section, we characterize our LAHs by two methods. 
One is to use the cumulative radial profiles of Ly$\alpha$ luminosity $L_{\text{Ly$\alpha$}}$ and rest-frame equivalent width $EW_0$ that allow us to investigate the details of the profiles with relatively small systematic uncertainties.
The other is to characterize the structure of our LAHs with the scale lengths $r_n$ (Section \ref{sec:sd}) that are quantities easily compared with the other physical parameters.

\subsection{Cumulative Radial Profiles of Ly$\alpha$ Luminosity and Equivalent Widths}
\label{sec:cum}
We derive cumulative radial profiles of $L_{\text{Ly$\alpha$}}$ and $EW_0$.
First, we calculate $L_{\text{Ly$\alpha$}}$ from the composite Ly$\alpha$ images.
Similarly, $EW_0$ is estimated from the composite Ly$\alpha$ and UV-continuum images.
The $EW_0$ value is defined as a ratio of $L_{\text{Ly$\alpha$}}$ to UV fluxes. 
Both $L_{\text{Ly$\alpha$}}$ and $EW_0$ cumulative radial profiles are obtained in $r=0-10''$ 
with a radius step of $0''.3$.
Figures \ref{fig:f_Lya} and \ref{fig:f_ew} show the cumulative radial profiles of our
subsamples.
Here we normalize $L_{\text{Ly$\alpha$}}$ and $EW_0$ cumulative radial profiles at $r=1''$
whose fluxes correspond to the $2''$-diameter aperture photometry.
The normalized $L_{\text{Ly$\alpha$}}$ and $EW_0$ cumulative radial profiles are labeled as $R$($L_{\text{Ly}\alpha}$) $_{r/1''}$  and $R$($EW_0$) $_{r/1''}$, respectively.
In Figures \ref{fig:f_Lya} and \ref{fig:f_ew}, we plot $R$($L_{\text{Ly}\alpha}$) $_{r/1''}$  and $R$($EW_0$) $_{r/1''}$ profiles up to a radius where the cumulative profile reaches the maximum value for clarity.

\subsubsection{$\log{L_{\text{Ly$\alpha$}}}$ Subsamples}
\label{sec:cmu_LLya}
\begin{figure*}
	\begin{center}
		\includegraphics[width=\linewidth]{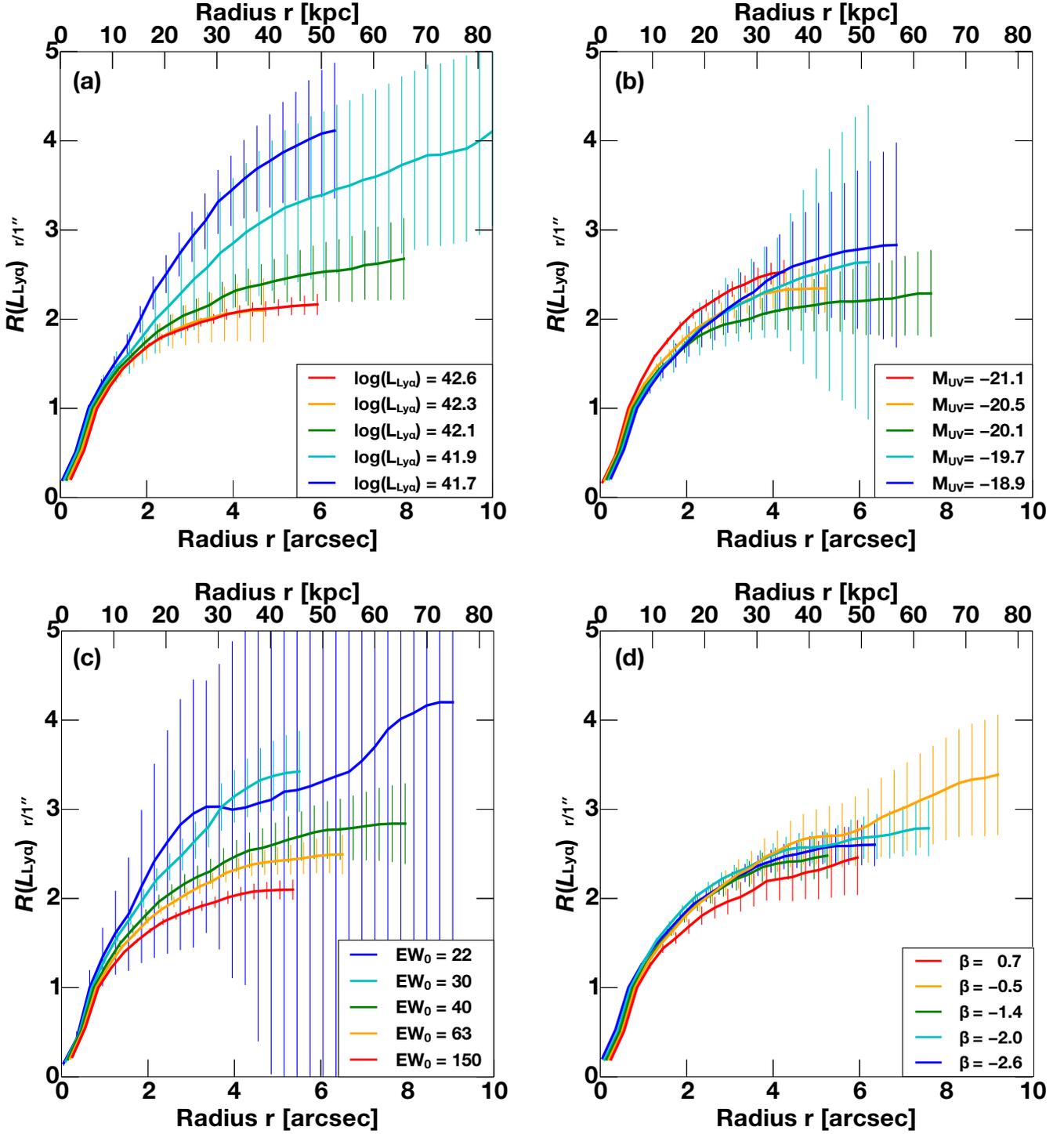} 
	\end{center}
	\caption{
	Cumulative radial profiles of $L_{\text{Ly}\alpha}$ normalized at $r=1''$.
	The panels of (a), (b), (c), and (d) present the subsamples of
	$\log{L_{\text{Ly$\alpha$}}}$, $M_{\text{UV}}$, $EW_0$, and $\beta$, respectively.
	See the labels in each panel for the line colors and the subsample names. 
	For clarity, we shift these profiles by $-0.1$, $-0.05$, $0$, $0.05$, and $0.1$ arcsec along the abscissa axis.
	}
	\label{fig:f_Lya}
\end{figure*}

The $R$($L_{\text{Ly}\alpha}$) $_{r/1''}$ profiles of the $\log{L_{\text{Ly$\alpha$}}}$ subsamples are presented in Figure \ref{fig:f_Lya} (a). 
There is a trend that the profiles of the $\log{L_{\text{Ly$\alpha$}}}$-faint subsamples are steeper than those of the $\log{L_{\text{Ly$\alpha$}}}$-bright subsamples at $r\gtrsim 10$ kpc.
At $r\simeq 40$kpc, $R$($L_{\text{Ly}\alpha}$) $_{r/1''}$ values are about 2 and 4 for the faint and bright subsamples of $\log{L_{\text{Ly$\alpha$}}}=42.6$ and $41.7$, respectively. In other words, the contribution to a total Ly$\alpha$ flux from the $r\gtrsim 10-40$ kpc range is about 3 ($=[4-1]/[2-1]$) times more for the Ly$\alpha$ faint ($\log{L_{\text{Ly$\alpha$}}}=41.7$) subsample than the Ly$\alpha$ bright ($\log{L_{\text{Ly$\alpha$}}}=42.6$) one, where the Ly$\alpha$ luminosities from the outskirt of the PSF are included.
It indicates that a significantly large fraction of Ly$\alpha$ luminosity
is emitted at the scale larger than $\sim 10$ kpc for the faint subsample.

\begin{figure*}
	\begin{center}
		\includegraphics[width=\linewidth]{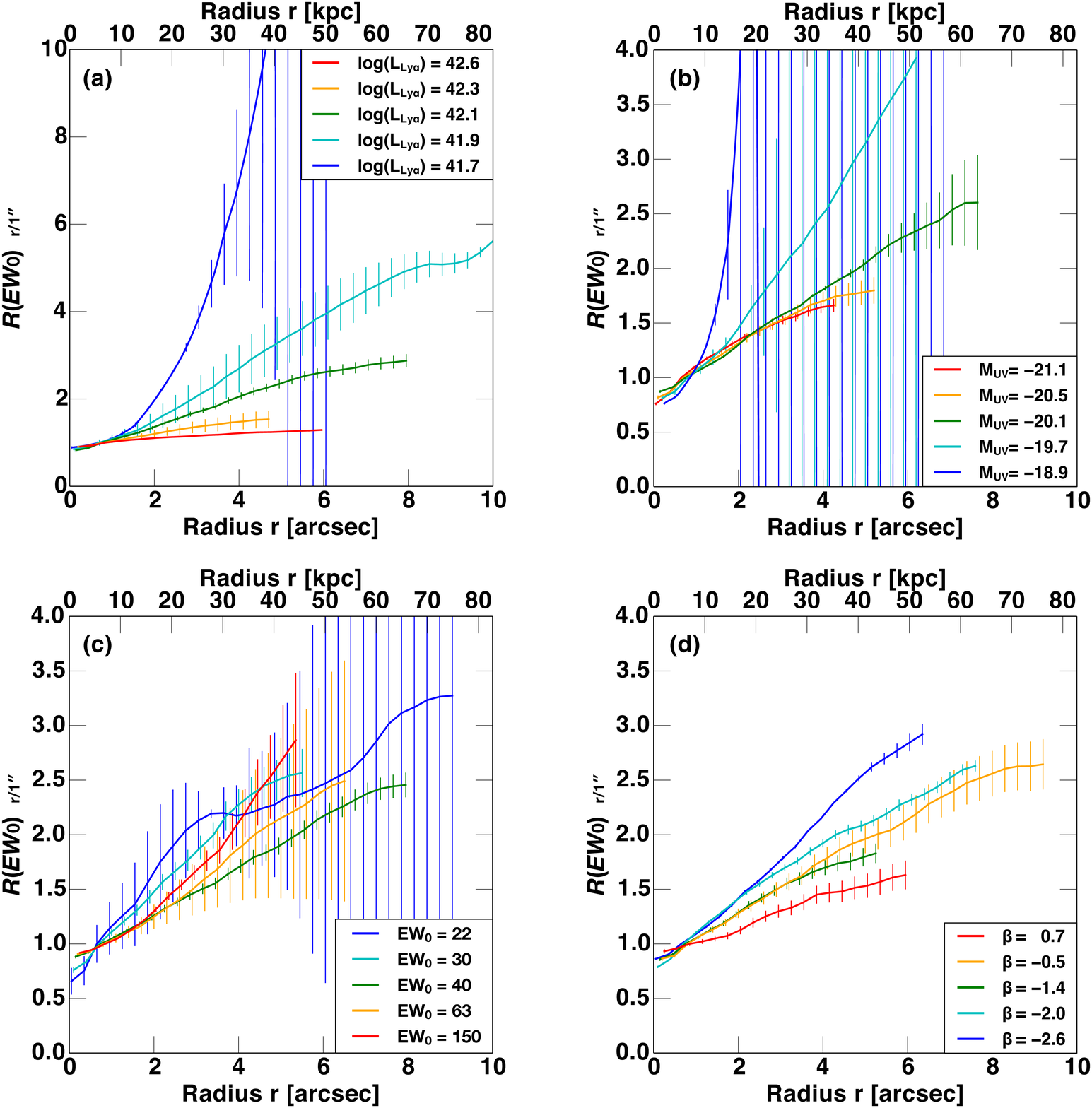} 
	\end{center}
	\caption{
	Same as Figure \ref{fig:f_Lya}, but for $EW_0$(Ly$\alpha$).	
	}
	\label{fig:f_ew}
\end{figure*}

Figure \ref{fig:f_ew} (a) presents $R$($EW_0$) $_{r/1''}$ profiles of the $\log{L_{\text{Ly$\alpha$}}}$ subsamples, and indicates that the increase of the Ly$\alpha$ SB is faster than the UV-continuum SB for any subsamples.
Similar to the $R$($L_{\text{Ly}\alpha}$) $_{r/1''}$ plot, in Figure \ref{fig:f_ew} (a) the $\log{L_{\text{Ly$\alpha$}}}$-faint subsamples show $R$($EW_0$) $_{r/1''}$ profiles steeper than the $\log{L_{\text{Ly$\alpha$}}}$-bright subsamples at $r\gtrsim 10$ kpc. This trend is consistent with the one of $R$($L_{\text{Ly}\alpha}$) $_{r/1''}$.

\subsubsection{$M_{\text{UV}}$ Subsamples}
In Figure \ref{fig:f_Lya} (b), we present $R$($L_{\text{Ly}\alpha}$) $_{r/1''}$ profiles of $M_{\text{UV}}$-subsamples.
Because two $M_{\text{UV}}$-faint subsamples of $M_{\text{UV}}=-19.7$ and $-18.9$ show very large uncertainties, the profiles of these two subsamples are indistinguishable from the other $M_{\text{UV}}$ subsamples at a large scale.
For the other three subsamples, $M_{\text{UV}}=-21.1$, $-20.5$ and $-20.1$, we find that the brighter $M_{\text{UV}}$ subsamples have the larger $R$($L_{\text{Ly}\alpha}$) $_{r/1''}$ values at $10-40$ kpc. This suggests that bright LAHs are associated with bright $M_{\text{UV}}$ sources that have a bright UV luminosity in a central $r=1''$ area.

Figure \ref{fig:f_ew} (b) presents $R$($EW_0$) $_{r/1''}$ profiles of $M_{\text{UV}}$-subsamples.
Similar to Figure \ref{fig:f_Lya} (b), the uncertainties of two faint subsamples ($M_{\text{UV}}=-19.7$ and $-18.9$) are too large to identify a trend. The three $M_{\text{UV}}$ subsamples ($M_{\text{UV}}=-21.1$, $-20.5$ and $-20.1$) have reasonably small error bars, and indicate that $R$($EW_0$) $_{r/1''}$ profiles of these three subsamples are comparable within the uncertainties.

\subsubsection{$EW_0$ Subsamples}
In Figrue \ref{fig:f_Lya} (c$)$, we present $R$($L_{\text{Ly}\alpha}$) $_{r/1''}$
profiles of the $EW_0$-subsamples. 
It is difficult to identify a difference of the $R$($L_{\text{Ly}\alpha}$) $_{r/1''}$ profiles between the $EW_0=22$ and the other $EW_0$-subsamples, due to large uncertainties in the profile of $EW_0=22$. 
For the other four subsamples we find a trend at $r\gtrsim10$ kpc that the $R$($L_{\text{Ly}\alpha}$) $_{r/1''}$ values are larger in the $EW_0$-smaller samples.
It indicates that a faint LAH is associated with an $EW_0$-large LAE.

The $R$($EW_0$) $_{r/1''}$ profiles of the $EW_0$-subsamples are shown in Figure \ref{fig:f_ew} (c$)$.
At $r\gtrsim10$ kpc, the $R$($EW_0$) $_{r/1''}$ profiles of the $EW_0=22$ and $63$ have large uncertainties that do not allow us to investigate the correlation between the profiles and the subsamples.
Although the other three $EW_0$-subsamples have small error bars, no clear trends of the profiles and the subsamples are found at $r\gtrsim10$ kpc.

\subsubsection{$\beta$ Subsamples}
\label{sec:cmu_beta}
The $R$($L_{\text{Ly}\alpha}$) $_{r/1''}$ profiles of the $\beta$-subsamples are presented in Figure \ref{fig:f_Lya} (d). 
These $R$($L_{\text{Ly}\alpha}$) $_{r/1''}$ profiles are similar within $1\sigma$ uncertainties.
There is no clear correlation between the $R$($L_{\text{Ly}\alpha}$) $_{r/1''}$ profiles and the LAEs' $\beta$ values defined in a central $r=1''$ area. 

We present the $R$($EW_0$) $_{r/1''}$ profiles of the $\beta$-subsamples in Figure \ref{fig:f_ew} (d).
In contrast to Figure \ref{fig:f_Lya} (d), we identify a significant difference of the $R$($EW_0$) $_{r/1''}$ profiles between the $\beta$-subsamples at a scale of $r\gtrsim10$ kpc.
There is a trend that the $R$($EW_0$) $_{r/1''}$ values at $r\gtrsim10$ kpc increase as the $\beta$ values of LAEs become small. 
In other words, an LAE with a blue UV continuum defined in a central $r=1''$ area has an LAH's $EW_0$ larger than those with a red UV continuum.

\subsubsection{Summary of the Trends Found in the Cumulative Radial Profiles}
We summarize the trends found in the cumulative radial profiles at $r\gtrsim10$ kpc presented in Section \ref{sec:cmu_LLya}$-$\ref{sec:cmu_beta}.
In Figures \ref{fig:f_Lya} (a), (b), and (c), the large $R$($L_{\text{Ly}\alpha}$) $_{r/1''}$ values are identified in the $\log{L_{\text{Ly$\alpha$}}}$-faint, $M_{\text{UV}}$-bright, and $EW_0$-small subsamples, respectively. 
The $\log{L_{\text{Ly$\alpha$}}}$-faint subsamples also have large $R$($EW_0$) $_{r/1''}$ values in Figure \ref{fig:f_ew} (a).
Our findings in Figures \ref{fig:f_Lya} (a)$-$(c$)$ and \ref{fig:f_ew} (a) indicate that LAEs with a faint $L_{\text{Ly}\alpha}$ luminosity, a bright UV luminosity, and/or a small $EW_0$ in a central $r=1''$ area possess prominent LAHs. In other words, galaxies with properties similar to LBGs, a faint $L_{\text{Ly}\alpha}$ luminosity, a bright UV luminosity, and a small $EW_0$ have strong LAHs.
We also identify the trend that the $\beta$-small subsamples have large $R$($EW_0$) $_{r/1''}$ values in Figure \ref{fig:f_ew} (d).

\subsection{Scale Lengths}
\label{sec:structure_param}

\subsubsection{Scale Length Measurements}
We estimate scale length $r_n$ values from the differential radial profiles in Figures \ref{fig:radi_all_1}$-$\ref{fig:radi_all_2}. The $r_n$ values are derived by the fitting of Equation \ref{eq:exp} to these profiles in the range of $r=2''$ to $40$ kpc in the same manner as Section \ref{sec:sd}. The best-fit $r_n$ values are summarized in Table \ref{tab:sample}.

\subsubsection{Correlations Between the Scale Lengths and the Other Physical Quantities}
\label{sec:results_rn}
\begin{figure}
	\begin{center}
		\includegraphics[scale=0.45]{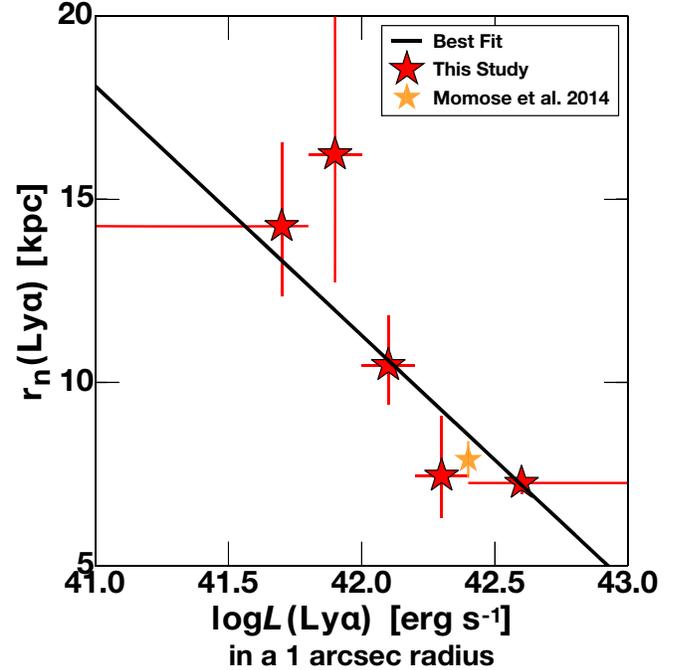} 
	\end{center}
	\caption{Scale length $r_n$ as a function of Ly$\alpha$ luminosity.
	The Ly$\alpha$ luminosity is measured in a $1''$-radius aperture.
	The red and orange stars represent the scale lengths estimated in this study and MO14, respectively.
	For the Ly$\alpha$ luminosities, we use the median $\log{L_{\text{Ly$\alpha$}}}$ values of the subsamples (see Table \ref{tab:sample}).
	The black solid line is the best-fit linear regression model of all the data.
	The vertical error bars are $1\sigma$ uncertainties of $r_n$ estimates, while horizontal error bars are the $\log{L_{\text{Ly$\alpha$}}}$ ranges of each subsample.
	}
	\label{fig:size_LLya}
\end{figure}

\begin{figure}
	\begin{center}
		\includegraphics[scale=0.45]{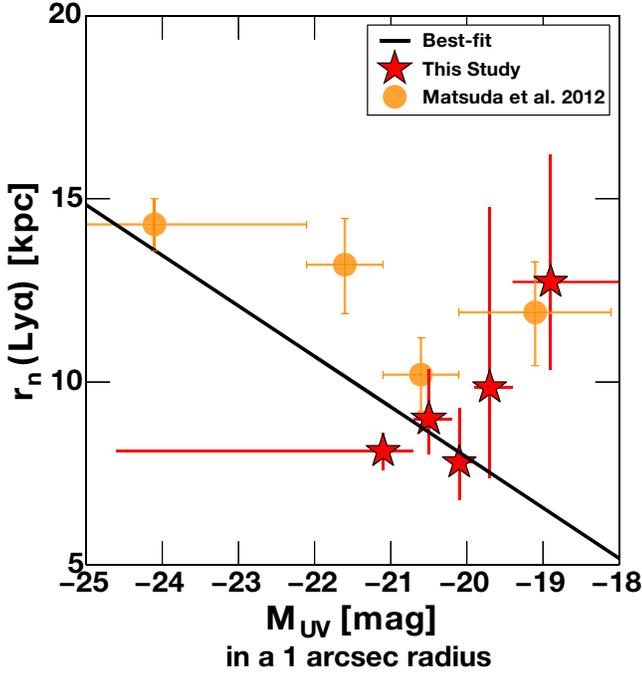} 
	\end{center}
	\caption{Same as Figure \ref{fig:size_LLya}, but for $M_{\text{UV}}$.
	The orange circles represent the $r_n$ estimates of MA12.
	The vertical error bars are $1\sigma$ uncertainties of $r_n$ estimates, while horizontal error bars are the $M_{\text{UV}}$ ranges of each subsample.
	}
	\label{fig:size_muv}
\end{figure}

\begin{figure}
	\begin{center}
		\includegraphics[scale=0.45]{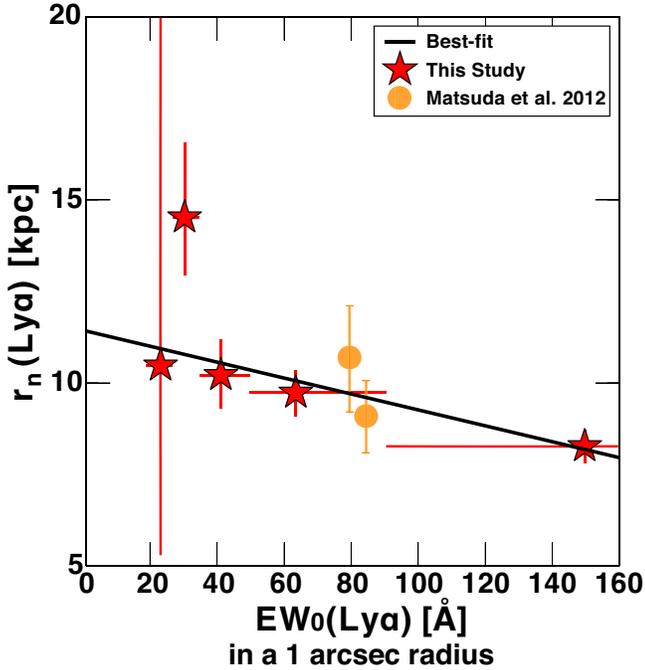} 
	\end{center}
	\caption{Same as Figure \ref{fig:size_LLya}, but for $EW_0$.
	See also the caption of Figure \ref{fig:size_LLya} for the legend symbols. 
	The vertical error bars are $1\sigma$ uncertainties of $r_n$ estimates, while horizontal error bars are the $EW_0$ ranges of each subsample.
	}
	\label{fig:size_ew}
\end{figure}

\begin{figure}
	\begin{center}
		\includegraphics[scale=0.45]{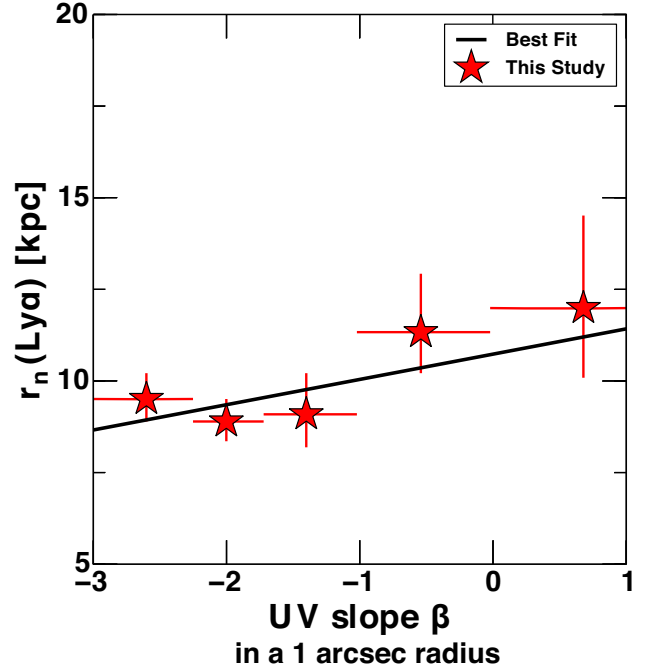} 
	\end{center}
	\caption{Same as Figure \ref{fig:size_LLya}, but for $\beta$.
	The vertical error bars are $1\sigma$ uncertainties of $r_n$ estimates, while horizontal error bars are the $\beta$ ranges of each subsample.
	}
	\label{fig:size_beta}
\end{figure}

We present the best-fit $r_n$ values as a function of the subsample median quantities of $\log{L_{\text{Ly$\alpha$}}}$, $M_{\text{UV}}$, $EW_0$, and $\beta$ in Figures \ref{fig:size_LLya}, \ref{fig:size_muv}, \ref{fig:size_ew}, and \ref{fig:size_beta}, respectively (see Table \ref{tab:sample}).  
We carry out the linear regression analysis to all the data presented in Figures \ref{fig:size_LLya}$-$\ref{fig:size_beta} that consist of our results and the previous study (MA12) measurements, if available.
The linear regression is evaluated by weighted least-squares fitting. 
The best-fit linear regression models are:
\begin{equation}\label{eq:rn_LLya}
	r_n = 297.1 - 6.8 \times \log{L_{\text{Ly$\alpha$}}},
\end{equation}
\begin{equation}\label{eq:rn_muv}
	r_n = -19.6 -1.4 \times M_{\text{UV}},
\end{equation}
\begin{equation}\label{eq:rn_ew}
	r_n = 11.4 -0.02 \times EW_0,
\end{equation} 
and 
\begin{equation}\label{eq:rn_beta}
	r_n = 10.7 + 0.7 \times \beta,
\end{equation}
that are shown with the black solid lines in Figures \ref{fig:size_LLya}$-$\ref{fig:size_beta}.
Figures \ref{fig:size_LLya}$-$\ref{fig:size_ew} and Equations \ref{eq:rn_LLya}$-$\ref{eq:rn_ew} show the anti-correlation between $r_n$ and a physical quantity of $\log{L_{\text{Ly$\alpha$}}}$, $M_{\text{UV}}$, or $EW_0$.
We should note that the anti-correlation between $r_n$ and $M_{\text{UV}}$, or $EW_0$ is weak, due to the large uncertainties of the $r_n$ estimates.
On the other hand, Figure \ref{fig:size_beta} and Equation \ref{eq:rn_beta} indicate 
a positive correlation between $r_n$ and $\beta$.

We calculate Spearman's rank correlation coefficients $\rho$ to evaluate the strengths of correlations between $r_n$ and the physical quantity ($\log{L_{\text{Ly$\alpha$}}}$, $M_{\text{UV}}$, $EW_0$, or $\beta$), and present the results in Table \ref{tab:param}.
The Spearman's $\rho$ values are $\rho=-0.9$ and $-0.7$ at $96\%$ and $93\%$ confidence levels for the correlations of $\log{L_{\text{Ly$\alpha$}}}$ and $EW_0$, respectively.
We thus find strong correlations to $\log{L_{\text{Ly$\alpha$}}}$ and $EW_0$ in the Spearman's $\rho$ estimates.
We also find the relatively strong correlation of $\beta$ whose Spearman's $\rho$ value is $\rho=0.7$ at an $81\%$ confidence level.
In contrast, the correlation of $M_{\text{UV}}$ is weak with the Spearman's $\rho$ of $-0.2$ ($45\%$ confidence level).
This is consistent with the result of Figure \ref{fig:size_muv} that the correlation between $r_n$ and $M_{\text{UV}}$ is not very clear, as discussed above.

\subsubsection{Summary of the $r_n$ Correlations}
In Section \ref{sec:results_rn}, $r_n$ shows anti-correlations with $\log{L_{\text{Ly$\alpha$}}}$, $M_{\text{UV}}$, and $EW_0$. There also exists a positive correlation between $r_n$ and $\beta$ (Figure \ref{fig:size_ew}). These correlations are almost consistent with the trends indicated in the cumulative radial profiles (Section \ref{sec:cum}), although the correlation between $r_n$ and $M_{\text{UV}}$ is not very clear.
In fact, MA12 identify no correlation between $r_n$ and $M_{\text{UV}}$ in their LAE samples. 
However, \citet{feld13} have reported that a UV brighter sample shows a more extended LAH, which is consistent with the anti-correlation between $r_n$ and $M_{\text{UV}}$ that is hinted in our study.

\begin{table}
\centering
\begin{minipage}{80mm}
\begin{center}
\caption{Results of the Correlation Tests and the Best-Fit Parameters}
\label{tab:param}
\begin{tabular}{cccccc}
\hline
(1)					& (2)		& (3)		& (4)		& (5)		& (6)		\\	
Quantity				& $\rho$	& $p_s$	& $R$	& $a$	& $b$	\\
\hline
$\log L_{\text{Ly$\alpha$}}$ & $-0.9$	& $0.04$	& $-0.88$	& $297.1$ & $-6.8$	\\
$M_{\text{UV}}^\dagger$	& $-0.2$	& $0.55$	& $-0.37$	& $-19.6$ 	& $-1.4$	\\ 
$EW_0^\dagger$		& $-0.7$	& $0.07$	& $-0.66$	& $11.4$ 	& $-0.02$	\\ 
$\beta$				& $0.7$	& $0.19$	& $0.88$	& $10.7$	& $0.7$	\\ 
\hline
\end{tabular}
\end{center}
The results are obtained from all the data consisting of our and MA12's data.
(1) Physical quantity for the correlation with $r_n$; 
(2) Spearman's rank correlation coefficient; 
(3) $p$-value of the Spearman's rank correlation test;
(4) Correlation coefficient;
(5) Best-fit intercept obtained from the linear regression model; 
(6) Best-fit slope obtained from the linear regression model. 
\end{minipage}
\end{table}


\section{Discussions}
\label{discussions}

\begin{figure}
	\begin{center}
		\includegraphics[width=\linewidth]{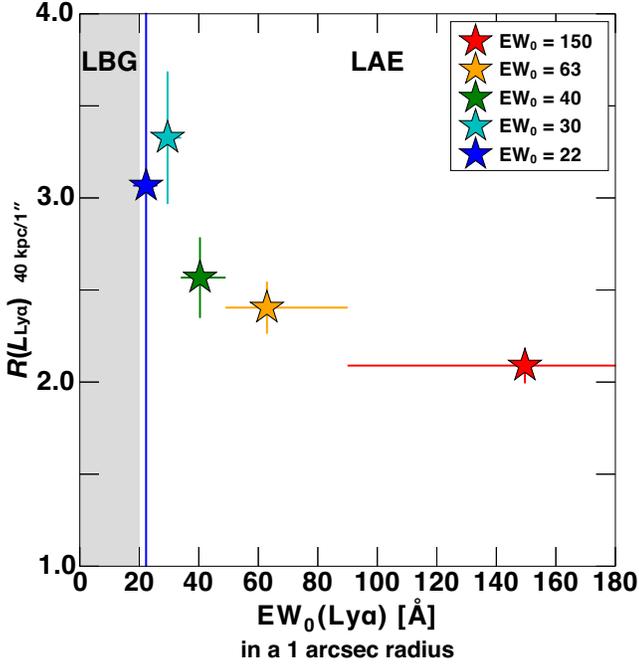} 
	\end{center}
	\caption{
	Ly$\alpha$ luminosity within $r=40$ kpc that is normalized by the one within $r=1''$.
	These normalized Ly$\alpha$ luminosities are shown as a function of
	$EW_0$ defined in a $r=1''$ area.
	The $EW_0$ range of each subsample is shown as horizontal error bars.
	}
	\label{fig:fracLyaewew}
\end{figure}

\subsection{Ly$\alpha$ Emission from the Diffuse LAHs}
\label{sec:lah_depend}
In Section \ref{sec:results}, both the cumulative radial profiles and the $r_n$ correlations clearly suggest that
the $EW_0$-small and $\log{L_{\text{Ly$\alpha$}}}$-faint LAEs have prominent LAHs.
Moreover, we find the extended LAHs at least up to $r\simeq 40$ kpc which corresponds to $r\simeq 4.''8$ 
in our composite Ly$\alpha$ images (Figures \ref{fig:image}$-$\ref{fig:radi_all_2}). 
Here we quantify the Ly$\alpha$ luminosities from LAHs within $r=40$ kpc 
as a function of $EW_0$ and $\log{L_{\text{Ly$\alpha$}}}$.

We plot the ratios of Ly$\alpha$ luminosities within $r=40$ kpc to those within $r=1''$ 
as a function of $EW_0$ in Figure \ref{fig:fracLyaewew}.
For simplicity, we refer to the ratios as $R$($L_{\text{Ly}\alpha}$) $_{40\text{kpc}/1''}$.
A large $R$($L_{\text{Ly}\alpha}$) $_{40\text{kpc}/1''}$ indicates more Ly$\alpha$ emission from LAHs.
Figure \ref{fig:fracLyaewew} shows the anti-correlation between $R$($L_{\text{Ly}\alpha}$) $_{40\text{kpc}/1''}$ and $EW_0$.
The $R$($L_{\text{Ly}\alpha}$) $_{40\text{kpc}/1''}$ value is high, about three for the LAEs with $EW_0\simeq 20$\AA. Because galaxies with $EW_0\lesssim 20$\AA\ dominate a sample of LBGs \citep{shapley2003}, Figure \ref{fig:fracLyaewew} suggests that LBG-like sources have LAHs whose Ly$\alpha$ luminosities are brighter than typical LAEs.
 
\begin{figure}
	\begin{center}
		\includegraphics[width=\linewidth]{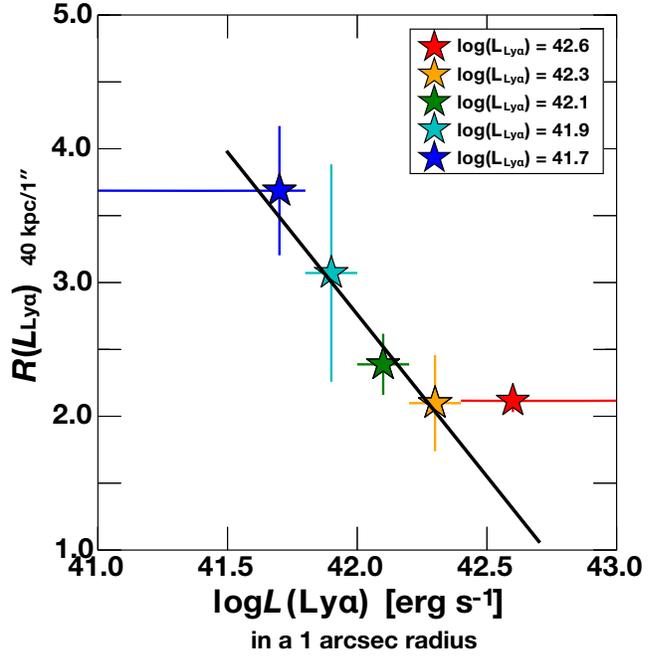}
	\end{center}
	\caption{Ly$\alpha$ luminosity within $r=40$ kpc that is normalized by the one within $r=1''$.
	These normalized Ly$\alpha$ luminosities are shown as a function of
	$EW_0$ defined in a $r=1''$ area.
	The gray shade represents the area that is bracketed by the two best-estimate linear functions of $L_{\text{Ly}\alpha}$($40$ kpc/$1''$) on $\log{L_{\text{Ly$\alpha$}}}$ and 
	$\log{L_{\text{Ly$\alpha$}}}$ on $L_{\text{Ly}\alpha}$($40$ kpc/$1''$). 
	The $\log{L_{\text{Ly$\alpha$}}}$ range of each subsample is shown as horizontal error bars.
	}
	\label{fig:fracLyaLya}
\end{figure}

Figure \ref{fig:fracLyaLya} presents $R$($L_{\text{Ly}\alpha}$) $_{40\text{kpc}/1''}$ as a function of $\log{L_{\text{Ly$\alpha$}}}$. 
Again, the $R$($L_{\text{Ly}\alpha}$) $_{40\text{kpc}/1''}$ value is high, about $2-4$, for the LAEs with $\log{L_{\text{Ly$\alpha$}}}\simeq 41.5-42.5$ erg s$^{-1}$, and there is a clear anti-correlation between $R$($L_{\text{Ly}\alpha}$) $_{40\text{kpc}/1''}$ and $\log{L_{\text{Ly$\alpha$}}}$.
We fit a linear function to Figure \ref{fig:fracLyaLya}, and obtain the best-fit function of 
\begin{equation}
\label{eq:lf}
	R(L_{\text{Ly}\alpha}) \ _{40\text{kpc}/1''} = 104.6 -2.4 \times \log{L_{\text{Ly$\alpha$}}} . 
\end{equation}
at $41.5$ erg s$^{-1}<\log{L_{\text{Ly$\alpha$}}}<42.8$ erg s$^{-1}$. 
In many previous studies, a $2-3''$-diameter ($\simeq1''$-radius) aperture magnitude or SExtractor \verb|MAG_AUTO| have been used to evaluate total Ly$\alpha$ luminosities and rest-frame Ly$\alpha$ equivalent widths of LAEs (e.g. \citealp{nil09,ouchi10,fin11a,naka12}). 
If the physical origin of LAHs is not external sources such as satellites (Section \ref{sec:satellite}), total Ly$\alpha$ luminosities may be underestimated in the previous studies with the amount indicated in Equation \ref{eq:lf}. In this case, the total-Ly$\alpha$ luminosity functions and densities may be revised.

\subsection{What is the Physical Origin of LAHs?}
\begin{figure*}
	\begin{center}
		\includegraphics[width=\linewidth]{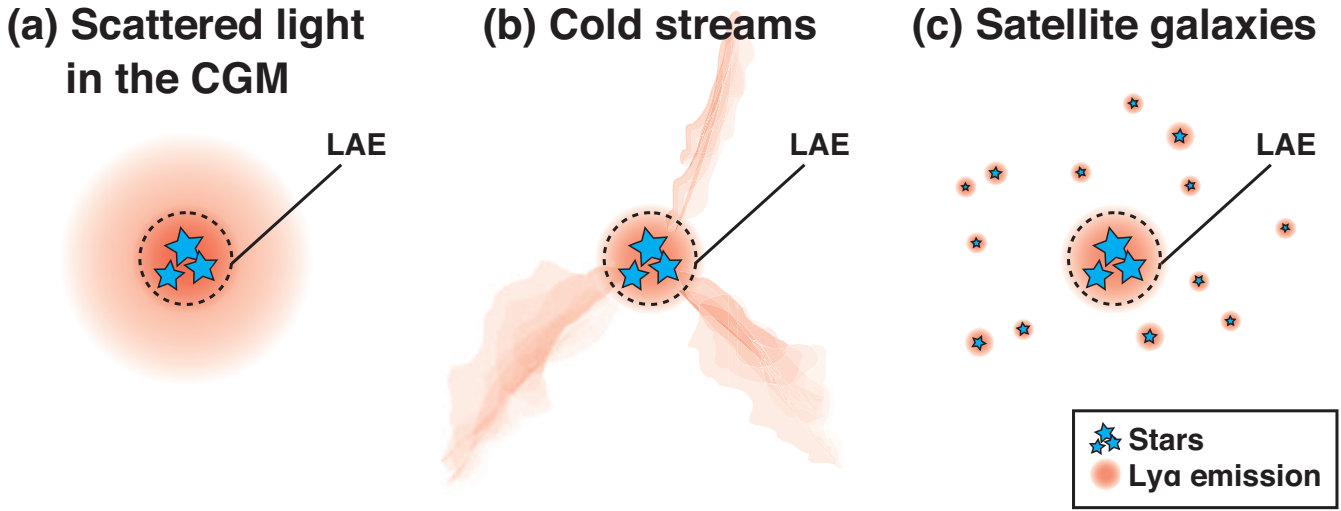} 
	\end{center}
	\caption{Illustrations of three possible origins of the LAHs,
	(a) scattered light in the CGM, (b) cold streams, and $($c) satellite galaxies.
	The cyan stars represent star-forming regions in the ISM.
	The red shades show ISM and CGM gas emitting or scattering Ly$\alpha$ that reaches the observer. The dotted circles denote the central regions of LAEs that are detected by observations on the individual basis.
	}
	\label{fig:scenario}
\end{figure*}

Theoretical studies suggest three physical origins of LAHs: (1) scattered light of H\,{\sc i} gas in the CGM, (2) cold streams, and (3) satellite galaxies. These three possible origins are illustrated in Figure \ref{fig:scenario}. In the following subsections, we discuss these possibilities with our findings in conjunction with
recent observation and simulation results.

\subsubsection{Scattered Light in the CGM}
\label{sec:scatter}
The first scenario is the scattered light of H\,{\sc i} gas in the CGM (Figure \ref{fig:scenario}a). 
In this scenario, Ly$\alpha$ photons are produced in star-forming regions and/or AGNs, and these Ly$\alpha$ photons escape from the interstellar medium (ISM) to the CGM.
The Ly$\alpha$ escape mechanism is key, but poorly understood.
Theoretical studies have proposed various mechanisms, such as outflows, clumpy clouds, and low column density of neutral hydrogen in the ISM $N_{\text{H\,{\sc i}, ISM}}$ (e.g. \citealp{neu91,verha06,hansen06,zhen11,orsi12,dijk12,duva14}). 
Because our study investigates neither spectra nor gas distribution of the ISM scale, no results from our study test this scenario.
Recent spectroscopic observations have reported the evidence of the outflow, although the velocity is as small as $\sim 200$ km s$^{-1}$ (e.g. \citealp{chon13,hashimoto13,shib14b}). 
\citet{scar09} have argued that their observational results could be reproduced by the clumpy dust distribution model of the ISM (see also \citealt{atek09,fin11b}).
Deep optical and near-infrared spectra for gas dynamics and line diagnostics indicate that the ISM of LAEs have a low $N_{\text{H\,{\sc i}, ISM}}$ (e.g. \citealp{hashimoto13,naka13,chon13,shib14a,shib14b,song14,naka14,par14}). 
So far, there are no conclusive observational tests including our results that rule out this first scenario.

\subsubsection{Cold Streams}
\label{sec:coldgas}
\begin{figure}
	\begin{center}
		\includegraphics[width=\linewidth]{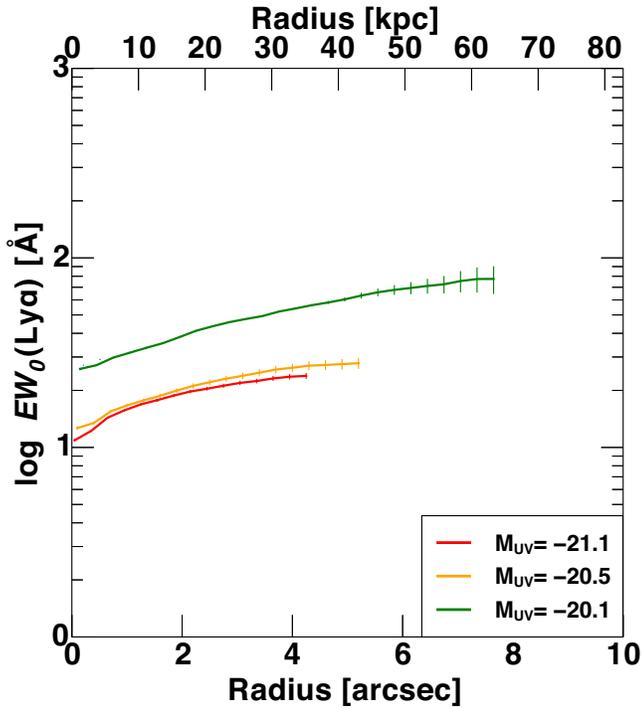} 
	\end{center}
	\caption{$EW_0$(Ly$\alpha$) as a function of radius for the $M_{\text{UV}}$ subsamples.
	The red, yellow, and green lines denote the subsamples of $M_{\text{UV}}=-21.1$, $-20.5$, and $-20.1$, respectively.
	Although two subsamples of $M_{\text{UV}}=-19.7$ and $-18.9$ reach $130$ and $1650$ {\AA} at the maximum, respectively, the error bars of two subsamples of $M_{\text{UV}}=-19.7$ and $-18.9$ are too large to distinguish between the large and small ($\lesssim 240$\AA) values in $EW_0$(Ly$\alpha$).
	We thus omit the data points of the $M_{\text{UV}}=-19.7$ and $-18.9$ subsamples for clarity.
	}
	\label{fig:cmu_muv}
\end{figure}

The second scenario is the cold streams (Figure \ref{fig:scenario}b). 
Cosmological hydrodynamical simulations suggest that intense star-formation of the high-$z$ galaxies ($z\sim 2$) is responsible for a dense and cold gas ($\sim10^4$ K) inflows that are dubbed cold streams (e.g. \citealp{keres05,dek09a,dek09b}). 
The cold streams radiate Ly$\alpha$ emission powered by gravitational energy, and produce an extended Ly$\alpha$ nebula around a galaxy.
Numerical simulations have indicated that a size of the cold stream Ly$\alpha$ nebula depends on a dark halo mass $M_{\text{DH}}$ \citep{ros12}.
Their massive ($M_{\text{DH}}\ge10^{12} M_\odot$) and less-massive ($M_{\text{DH}}\sim10^{11} M_\odot$) galaxies have large and small Ly$\alpha$ nebulae whose sizes are $\gtrsim 100$ and $\simeq 20$ kpc in radius, respectively, because more Ly$\alpha$ photons are produced in and around massive halos.
Since clustering analyses of high-$z$ galaxies show a positive correlation between dark halo masses and UV luminosities \citep{ouchi04b,lee06}, the dark halo mass dependence can be investigated with UV magnitudes.
Our results in Section \ref{sec:results} indicate that large LAHs are found in UV luminous LAEs.
These results are consistent with the scenario that the LAHs are produced by the cold streams.
However, Figure \ref{fig:cmu_muv} shows that $EW_0$(Ly$\alpha$) values of our $M_{\text{UV}}$ subsamples are lower than $77$ {\AA}. 
If the cold streams are responsible for the LAHs under the circumstances that the majority of Ly$\alpha$ photons generated in a central galaxy could avoid dust extinction and escape to the LAH, the $EW_0$(Ly$\alpha$) values at large radii should be larger than $240$ {\AA} that is the maximum value for Ly$\alpha$ photons originating from regular population II star formation \citep{mal02}.
It indicates that the majority of Ly$\alpha$ photons of our LAHs are not produced in the cold streams.

\subsubsection{Satellite Galaxies}
\label{sec:satellite}
The third scenario is the satellite galaxies (Figure \ref{fig:scenario}c). Satellite galaxies exist around an LAE, and radiate Ly$\alpha$ emission. If the total radiation from the satellite galaxies is strong, satellite galaxies would produce the extended Ly$\alpha$ emission structure around the LAE in the composite Ly$\alpha$ image. The extended Ly$\alpha$ emission structure may be identified as the LAH. 
Some cosmological simulations also indicate the presence of extended LAHs due to the Ly$\alpha$ emission from satellite galaxies \citep[e.g.][]{shim10,lake15}.
If there exists the significant Ly$\alpha$ contribution from satellite galaxies, an extended UV emission made by stellar components of the satellites would be shown in the composite UV-continuum images. We have found no such an extended UV emission in our composite UV-continuum images of Figures \ref{fig:image}$-$\ref{fig:radi_all_2}. 
However, our composite UV-continuum images suffer from sky over-subtraction at the level of $4\times10^{-33}$ erg s$^{-1}$ cm$^{-2}$ Hz$^{-1}$ arcsec$^{-2}$ as described in Section \ref{nonLAE_stacking}. It may cancel out the UV emission from satellite galaxies.
We thus cannot reach a conclusion about the satellite-galaxy contribution to the LAHs with our data.

\subsubsection{Remaining Possibilities for the Physical Origins of LAHs}
In Sections \ref{sec:scatter}$-$\ref{sec:satellite}, we discuss three possible physical origins of LAHs: 
(1) the scattered light in the CGM, 
(2) the cold streams, and 
(3) satellite galaxies. 
Our results rule out the possibility that (2) the cold streams is a major contributor to the LAHs. Because there remain the two possibilities of (1) the scattered light in the CGM and (3) satellite galaxies, we cannot conclude one most likely scenario.

Recently, theoretical studies have claimed that LAHs cannot be reproduced only by the scattered light in the CGM whose emission is originated from a central galaxy.
\citet{lake15} have used hydrodynamic simulations including galaxy formation, and shown a drop-off of the Ly$\alpha$ radial SB profile at $r\sim20$ kpc that does not agree with the results of MO14.
On the other hand, the simulations of \citet{lake15} reproduce the Ly$\alpha$ radial SB profile at $r>20$ kpc of the MO14's LAHs, if they include the contribution from satellite galaxies in their simulations.
If these simulation results are correct, the LAH at $r>20$ kpc would be produced by the satellite galaxies.

\section{Summary}
\label{summary}
In this paper, we investigate LAHs' radial SB profiles and scale lengths $r_n$ as a function of the four physical quantities of central galaxies, Ly$\alpha$ luminosity $\log L_{\text{Ly$\alpha$}}$, rest-frame UV magnitude $M_{\text{UV}}$, rest-frame Ly$\alpha$ equivalent width $EW_0$, and UV slope $\beta$, and discuss LAHs' properties and origins.
The major results of this paper are summarized below.

\begin{enumerate}
	\setlength{\leftskip}{0.5cm}
	\renewcommand{\theenumi}{\arabic{enumi}.}
	\item We detect LAHs from all LAE subsamples of $\log L_{\text{Ly$\alpha$}}$, $M_{\text{UV}}$, $EW_0$ and $\beta$, where we carefully examine the statistical and systematic errors in both the composite Ly$\alpha$ and UV images with the SB radial profiles of the non-LAE composite images. Comparing the non-LAE composite images with the LAE composite images, we rule out the possibility that a total of systematic uncertainties (i.e. artifacts) do not produce the LAHs found in our LAE composite data. We confirm that $r_n$ of Ly$\alpha$ radial SB profiles of our LAHs are consistent with those previously obtained in MA12 and MO14.
	\item We characterize our LAHs with the cumulative radial profiles of $L_{\text{Ly$\alpha$}}$ and $EW_0$. Here, the cumulative radial profiles are shown with $R$($L_{\text{Ly}\alpha}$) $_{r/1''}$ and $R$($EW_0$) $_{r/1''}$ that are defined as $L_{\text{Ly$\alpha$}}$ and $EW_0$ values within an aperture of radius $r$, respectively, which are normalized at $r=1''$.
	We find that the $R$($L_{\text{Ly}\alpha}$) $_{r/1''}$ values are large in the $\log L_{\text{Ly$\alpha$}}$-faint, $M_{\text{UV}}$-bright, and $EW_0$-small subsamples (Figure \ref{fig:f_Lya}a$-$c). Similarly, we also find the large $R$($EW_0$) $_{r/1''}$ values in the $\log L_{\text{Ly$\alpha$}}$-faint subsamples. These results indicate that there are prominent LAHs around LAEs that have a faint Ly$\alpha$ luminosity, a bright UV luminosity, and/or a small $EW_0$ (Figures \ref{fig:f_Lya}$-$\ref{fig:f_ew}). In other words, galaxies with properties similar to LBGs, a faint $L_{\text{Ly}\alpha}$ luminosity, a bright UV luminosity, and a small $EW_0$ have strong LAHs.
	\item We estimate $r_n$ values from the differential radial Ly$\alpha$ SB profiles of our LAHs, and investigate correlations between $r_n$ and four physical quantities of $\log L_{\text{Ly$\alpha$}}$, $M_{\text{UV}}$, $EW_0$, and $\beta$ (Figures \ref{fig:size_LLya}$-$\ref{fig:size_beta}). We find anti-correlations between $r_n$ and $\log L_{\text{Ly$\alpha$}}$, $M_{\text{UV}}$, or $EW_0$, and a positive correlation between $r_n$ and $\beta$. The Spearman's $\rho$ estimates suggest that there exist significant correlations in those of $\log L_{\text{Ly$\alpha$}}$, $EW_0$, and $\beta$, while the $\rho$ estimates indicate that the correlation between $r_n$ and $M_{\text{UV}}$ is not very clear.
	\item We identify a clear anti-correlation between $R$($L_{\text{Ly}\alpha}$) $_{40\text{kpc}/1''}$ and $\log{L_{\text{Ly$\alpha$}}}$, where $R$($L_{\text{Ly}\alpha}$) $_{40\text{kpc}/1''}$ is $R$($L_{\text{Ly}\alpha}$) $_{r/1''}$ for $r=40$ kpc (Figure \ref{fig:fracLyaLya}). The $R$($L_{\text{Ly}\alpha}$) $_{40\text{kpc}/1''}$ value is high, about $2-4$, for the LAEs with $\log{L_{\text{Ly$\alpha$}}}\simeq 41.5-42.5$ erg s$^{-1}$, and there is a clear anti-correlation between $R$($L_{\text{Ly}\alpha}$) $_{40\text{kpc}/1''}$ and $\log{L_{\text{Ly$\alpha$}}}$. 	If the majority of Ly$\alpha$ emission of LAHs are not originated from external sources such as satellites (Section \ref{sec:satellite}), total Ly$\alpha$ luminosities of high-$z$ galaxies may be underestimated in a popular $\simeq 1''$-radius aperture photometry 	by the amount indicated in Equation \ref{eq:lf}. If it is true, the total-Ly$\alpha$ luminosity functions 	and densities would be revised.
	\item With our results, we discuss three scenarios for the origin of LAHs : (1) the scattered light of H\,{\sc i} gas in the CGM, (2) the cold streams, and (3) satellite galaxies. Our cumulative radial profiles of rest-frame Ly$\alpha$ equivalent width do not support the cold stream scenario (2), because the Ly$\alpha$ equivalent width reaches only $77$\AA\ that is significantly smaller than 240\AA, the maximum value for Ly$\alpha$ photons originating from regular population II star formation. On the other hand, our results do not test the scenarios of (1) and (3). We thus conclude that there remain two possible scenarios of (1) and (3).
\end{enumerate}

\section*{Acknowledgements}

We thank Alex Hagen, Caryl Gronwall, Lucia Guaita, 
Yuichi Matsuda,
Michael Rauch, James Rhoads, Anne Verhamme, and Zheng Zheng
for useful comments and discussions.
This work was supported by World Premier International Research
Center Initiative (WPI Initiative), MEXT, Japan,
and KAKENHI (23244025) and (15H02064)
Grant-in-Aid for Scientific Research
(A) through Japan Society for the Promotion of Science
(JSPS). K.N. and S.Y. acknowledge the JSPS Research Fellowship
for Young Scientists.

{}


\end{document}